\documentclass[draft,showpacs,preprintnumbers,amsmath,amssymb,fourier]{revtex4-2}

\usepackage{xcolor}
\usepackage{soul}
\usepackage{amsmath, amsthm, amsfonts}
\usepackage{amssymb}

\theoremstyle{definition}

\theoremstyle{remark}

\begin{document}

\title{Gravitational waves in  massive Horndeski theory with a potential}

\begin{abstract}

We investigate gravitational waves with an arbitrary potential within the framework of linearized Horndeski theory. We show that the minimum of the potential can play the role of an effective cosmological constant in this theory, which is usually neglected in previous studies of this subject. We first determine the background geometry in this setup  by solving the weak field scalar and tensorial equations of linearized Horndeski theory. The solutions of linearized weak-field wave equations, in an appropriate gauge, are then obtained perturbatively to study the propagation and interactions of gravitational waves in this background. We compare our results with different realizations of the cosmological constant in Horndeski theory to compare the role of an arbitrary scalar potential with those of vacuum energy density and a linear potential. The results show that the background curvature arising from the minimum of the scalar potential effectively mimics a cosmological constant, producing distinct redshifts in the frequency and wave number that distinguish the tensor waves from massive scalar ones. We also find that the way the cosmological constant is introduced directly influences the speed and polarization of the scalar wave.

\end{abstract}

\pacs{04.30.-w,04.30.Nk,04.50.Kd}

\author{Hatice \"Ozer}
\email{hatice.ozer@istanbul.edu.tr (corresponding author)}
\affiliation{Department of Physics, Faculty of  Sciences,  Istanbul University, 34134  Istanbul, Türkiye\\
ORCID:0000-0002-6325-0598}
\author{\"Ozg\"ur Delice}
\email{ozgur.delice@marmara.edu.tr}
\affiliation{Department of Physics, Faculty of Science, Marmara University, 34722  Istanbul, Türkiye\\
ORCID: 0000-0001-9494-8617}
\date{\today}

\maketitle

\newpage 
\section{Introduction}

On September 14th, 2015,  the Laser Interferometer Gravitational-Wave Observatory (LIGO) Scientific Collaboration and the Virgo Collaboration formally announced that they had successfully detected the first gravitational waves (GWs) directly \cite{Abbott, Abbott1, Abbott2,Abbott3}. It is worth mentioning that this milestone was achieved almost a century after Einstein predicted the existence of GWs based on his theory of General Relativity (GR). The obtained results are consistent with the predictions of GR.
The discovery of gravitational waves by the LIGO not only marks the start of a new era called multi-messenger astronomy \cite{Ligomm}, but also aims at exploiting these waves for testing \cite{Yunes,Will,Will2} alternative theories of gravity.  Nevertheless, despite its remarkable success, GR still leaves  several unresolved fundamental problems in modern physics, such as the origin of cosmic acceleration \cite{Perlmutter, Riess,Riess1,SDSS}, the nature of dark matter and dark energy, and the absence of a consistent quantum theory of gravity \cite{Oriti}. These unresolved issues strongly motivate the study of alternative gravitational theories \cite{Sotiriou,Capozzielo,Nojiri1,Clifton,Nojiri2}. One of the most significant challenges faced by GR is explaining cosmic acceleration. While the accelerated expansion discovered in the late 1990s can be explained within GR by introducing a cosmological constant $\Lambda$, the extreme fine-tuning required for its value gives rise to the well-known cosmological constant problem \cite{AdlerCC}. To overcome the fine-tuning difficulties associated with $\Lambda$, scalar–tensor theories \cite{Jordan,Brans1} introduce a dynamical scalar field that can act as the source of cosmic acceleration. Particularly, Horndeski theory \cite{Horndeski} provides alternative explanations for both dark matter and dark energy \cite{Clifton}. Through the modification of gravitational interactions, they are able to generate dark matter–like effects on galactic and cosmological scales, while at the same time remaining promising candidates for addressing the dark energy problem through evolving scalar field dynamics \cite{Sotiriou,Capozzielo,Nojiri1,Clifton,Nojiri2}. In this context, scalar–tensor theories, and in particular the most general one having second-order field equations, known as the Horndeski theory, can play a central role in the exploration of modified gravity theories.  

Scalar-tensor theories \cite{Jordan,Brans1} represent the most basic alternative theory of gravity that involves a scalar field in addition to the metric tensor for the gravitational interaction. For a history of earlier developments of scalar-tensor theories, see \cite{Goenner}. In the basic form, they generalize GR to include such a scalar field, but later, more general scalar-tensor theories were presented by coupling a scalar field to higher-order curvature scalars. For a review of several aspects of modified or alternative theories of gravity, including recent developments of scalar-tensor theories, we refer to the latest reviews \cite{Sotiriou,Capozzielo,Nojiri1,Clifton,Nojiri2}. 
Horndeski theory \cite{Horndeski} is the most general scalar-tensor theory leading to at most second-order field equations.  This theory can be considered as a framework theory whose special cases include well-known important theories such as  GR, Brans-Dicke (BD), $f(R)$, quintessence theory, etc.  It is also one of the most important alternatives or generalizations to the general theory of relativity in this manner.  

Scalar fields have various applications and usages in gravity theories, either as a source or as a part of gravitational interaction via their couplings to the curvature in various scalar-tensor theories. 
For example, they are used in unification schemes such as string theory \cite{Lidsey},  and enable a model for cosmological inflation \cite{Starobinsky}.  A scalar field with a slowly changing potential could be a candidate for inflation \cite{Linde} as well as dark energy \cite{Copeland}. Besides, the cosmological constant ($\Lambda$) and the minimum value of the potential of scalar fields can be shown as the most important candidates for large-scale background curvature. In GR, the constant background curvature, $\Lambda$ corresponds to vacuum (dark) energy. However, this equivalence does not generally hold in alternative theories of gravity. For example, in GR the cosmological constant $\Lambda$ and the vacuum energy (modeled as a perfect fluid with equation of state $p=-\rho$ where $\rho=\Lambda$) are equivalent, whereas in the BD theory a $\Lambda$ term \cite{Lorenz-Petzold,Endo,Kim} and vacuum energy \cite{Mathiazhagen,La,Weinberg} behave differently as background curvature contributions, as shown in \cite{Barrow}. It was later shown in \cite{Ozer} that the minimum of a potential as a background curvature mimicking cosmological constant also behaves differently than cosmological constant in BD theory.

Another aspect of adding a scalar potential in scalar-tensor theories \cite{Bergman,Nordvedt,Wagoner} is that it generally leads to a massive theory \cite{Will2,Alsing,Ozer}, rendering the scalar field short-range. The resulting massive theory will have dramatically different local and asymptotic behavior. It shows screening effects \cite{Vainshtein,Khoury,Khoury1} where the scalar field frozen beyond a specific range determined by the mass of the scalar field. This is contrast to the original BD theory (with \cite{Ozer1} or without \cite{Brans1} $\Lambda$  or dark energy) where the scalar field is massless and has a long range. In summary, counterparts to  cosmological constant  in scalar-tensor theories have a rich structure with different local and global behavior and with associated short and long-range scalar fields. Hence, in BD theory, there are three different ways of treating the $\Lambda$. It can be considered as part of the spacetime curvature, introduced as a scalar field potential, or added to the energy–momentum tensor as vacuum energy. These different treatments affect the field equations in different ways and leads to different physical consequences. Therefore, the background curvature has distinct properties in the alternative theories of gravity.

This paper is devoted to studying the properties of gravitational waves (GWs) in linearized Horndeski gravity on asymptotically nonflat background geometries. GWs in linearized approximation, together with properties such as polarization and propagation speed have previously been studied for several sub-theories, including BD theory \cite{Wagoner,Will3,Alsing,Maggiore}, $F(R)$ gravity \cite{Capozziello, Corda,Naf,Berry,Kausar,Dicong}, and  generic linear massive theories \cite{Tachinami}. These analyses were mainly carried out in asymptotically flat backgrounds, including Horndeski theory itself \cite{Houa,Moretti}. Now, it is time to expand those works to the asymptotically nonflat background. Hence, our primary goal is to consider the properties of GWs and their propagation in the framework of Horndeski theory, taking into account the effect of the background curvature due to the potential of the scalar field, leading to an asymptotically nonflat de Sitter-type background.  Indeed, asymptotically nonflat linearized waves are mostly studied within the framework of GR on backgrounds with a cosmological constant $\Lambda$ \cite{NafJetzer,Bernabeu1,Espriu,Arraut,EspriuPTA} or a general cosmic background \cite{Espriu1,EspriuRodera}. This analysis is rarely extended to alternative gravity theories, except for a recent work \cite{Ozer1}. Hence, in this paper, we extend these analyses to  Horndeski theory. We consider a nonvanishing scalar potential acting as an effective cosmological constant, a linear potential leading to a cosmological constant with a massless scalar field, and a vacuum energy density filling the universe and  playing the role of $\Lambda$ as in GR. In particular, we will discuss the effects of the curvature of the spacetime background due to the potential of the Horndeski theory on the gravitational wave equations, their solutions, propagation, and polarization states to reveal cosmological effects on detectors, compared to a linear potential case or a cosmic vacuum dark energy fluid filling the spacetime.

Properties of GWs, such as their polarization properties, provide a new and powerful test for alternative theories of gravity. Also, the observation of polarization modes of GWs can be used to obtain important information about astronomical objects from which the waves are generated. Therefore, it is worthwhile to explore the polarization content of GWs of different scenarios. We know that general scalar-tensor-vector theories of gravity suggest that there can exist at most six possible polarization modes for GWs: two tensor polarization modes, two vector polarization modes, and two scalar polarization modes \cite{Will2, Callister, Thorne}. GR theory predicts that gravitational radiation is purely tensor polarized, called plus (+) and cross (×) tensor modes. Scalar-tensor theories and some theories with extra dimensions predict the existence of a breathing scalar mode. Our study delves into the polarization properties of GWs in massive Horndeski theory, where GWs contain a massive scalar mode in addition to the tensor modes. This theory predicts that, in addition to the well-known + and × polarizations, GWs contain a single massive scalar mode, which induces both transverse breathing and longitudinal components, forming a mixed polarization state \cite{Dicong,Houa}. Nevertheless, it is difficult to obtain information about the polarization content of such GWs using signals from the three-detector LIGO-Virgo network \cite{Abbott4,Abbot5,Chatziioannou, Isi}. For transient events observed by the LIGO-Virgo collaboration, the two LIGO detectors are nearly aligned and therefore have high sensitivity to only a single polarization mode.  Moreover, the three GW detectors of the LIGO-Virgo network are insufficient to fully resolve the complete polarization content of GWs, and at least five detectors are required for this purpose \cite{Abbot5,Chatziioannou,Isi}. Nevertheless, when considered together, the detectors can distinguish whether the observed waves are purely tensorial or purely scalar. In particular, for three-detector events, a purely tensorial signal is favored over purely scalar or purely vector signals \cite{Abbott4}. Moreover,  a single interferometer cannot detect the scalar breathing modes by design. All available observations of GWs are consistent with predictions from GR. Hence, the detection of new polarization modes through interferometers would be extremely challenging. However, future detectors \cite{Lisa} are designed to search for six different types of polarization and variations in the propagation speed of the waves. The long-duration GW sources such as stochastic background \cite{Nishizawa,Callister}, persistent signals from neutron stars \cite{isi1},  or the use of Pulsar Timing Arrays \cite{ipta,pta} might serve as a more effective tool to detect deviations from predictions of  GR, such as scalar or vector GWs.

This paper is organized as follows. In section II, we review the linearized field equations of the Horndeski theory in the presence of a non-trivial potential for an asymptotically non-flat geometry caused by the minimum of the potential. In section III,  we will obtain background and linearized gravitational wave solutions in linearized Horndeski theory in this setup. In section IV, we will discuss the propagation of GWs and their interaction with test masses or detectors using geodesic deviation equations. In section V, we will transform these solutions to the Friedmann–Lemaître–Robertson–Walker (FLRW) coordinates to better see the cosmological effects of the background curvature originating from the scalar potential in Horndeski theory on the gravitational waves. In section VI, we will obtain the solutions and compare their physical outcomes for the other approaches to have a cosmological constant, such as a linear potential case and a vacuum energy density case, in comparison with the arbitrary potential case discussed in sections II-V.
 The paper ends with a brief discussion. 

\section{Linearized Field Equations in Horndeski Theory with a Potential around a Minkowski Background}

Horndeski theory is the most general scalar-tensor theory of gravity having second-order field equations in four dimensions. The action of the Horndeski theory is defined as follows \cite{Horndeski},
\begin{equation}\label{Hornaction}
	S= \int d^4x \sqrt{-g}(L_2+L_3+L_4+L_5)+ \int dx^4 \sqrt{-g} L_m,
\end{equation}
where
\begin{align}
	L_2&= K(\phi,X), \quad L_3= -G_3(\phi, X) \Box\phi, \nonumber \\
	L_4&= G_4(\phi,X) R + G_{4,X} \left[(\Box\phi)^2 - (\nabla_\mu\nabla_\nu\phi)(\nabla^\mu\nabla^\nu\phi)\right], \label{Hornactions} \\
	L_5&= G_5(\phi,X) G_{\mu\nu} \nabla^\mu\nabla^\nu\phi - \frac{1}{6}G_{5,X} \left[(\Box\phi)^3 
	- 3(\Box\phi)(\nabla_\mu\nabla_\nu\phi)(\nabla^\mu\nabla^\nu\phi) 
	+ 2(\nabla^\mu\nabla_\alpha\phi) (\nabla^\alpha\nabla_\beta\phi) (\nabla^\beta\nabla_\mu\phi) \right]. \nonumber
\end{align}  
Here $g$ is the determinant of the metric tensor, $\Box=g_{\mu\nu}\nabla^\mu\nabla^\nu$ is the D'Alembertian operator, 
the functions $K$, $G_3$, $G_4$, $G_5$ are arbitrary functions of $\phi$ and $X$, where   $X=-\nabla_\mu\phi\nabla^\mu\phi/2$ is a canonical kinetic term and $L_m$ is the matter-energy Lagrangian. Here we use the notations like $G_{j,X}(\phi,X)=\frac{\partial G_j(\phi,X)}{\partial X}$ with $j=4,5.$ 

In order to derive the weak field equations for GWs, we define the metric as follows \cite{Will2},
\begin{align}
	g_{\mu\nu} &= \eta_{\mu\nu} + h_{\mu\nu},\quad  g^{\mu\nu}=\eta^{\mu\nu}-h^{\mu\nu}, \\
	\phi &= \phi_0 + \varphi,
\end{align}
where $\eta_{\mu\nu}$ is the Minkowski metric, $\phi_0$ is a constant value of the scalar field, $h{_{\mu\nu}}$ is the metric perturbation tensor and $\varphi$ is the scalar perturbation with the conditions, $|h{_{\mu\nu}}|\ , \varphi\ll 1$. Our goal is to obtain the gravitational wave equations around the non-flat background, originating from the potential term possibly included in the term $K$. Hence, we will use the condition, $K\neq 0$.
We also need proper linearization of $K(\phi,X)$. In the linear approximation, we have to take $X=0$ since it is quadratic in $\varphi$. Under this assumption, we assume that there exists a potential term in $K(\phi,X)$ in which it does not vanish when we take $X=0$. Let us for clarity we denote $K(\phi,0)=K(\phi)$. Then we can expand this term in a Taylor series around $\phi=\phi_0$, giving
\begin{eqnarray}\label{Kexpansion}
	K(\phi)&=&K(0)+K_{,\phi}(0)\varphi+\frac{1}{2}K_{,\phi\phi}(0)\varphi^2+\ldots,\\
	K_{,\phi}(\phi)&=&K_{,\phi}(0)+K_{,\phi\phi}(0)\varphi+\ldots,\nonumber
\end{eqnarray}
where we use the shorthand $K(0)\equiv K(\phi_0,0)$, etc. It might be reasonable for stability to have $\phi_0$ to be the minimum of the potential term present in $K$. Hence we can assume that $K_{,\phi}(0)=0$.
We also suppose that, unlike the previous works on scalar-tensor theories \cite{Alsing,Houa,Berry} demanding asymptotic flatness, minimum of the potential and vacuum energy density, $\Lambda$ that we will use later, are first-order quantities, i. e., $\Lambda, K(0)\sim \mathcal{O}(\eta, h). $ This will ensure that the spacetime we are working with is an asymptotically de Sitter one. 

Linearized gravitational waves in Horndeski theory are studied before \cite{Houa}. In those works, the background curvature stemming from the potential is ignored to obtain an asymptotically flat background geometry. Hence, in those works, the conditions
\begin{equation}
	K(0)=0,\quad K_{,\phi}(0)=0
\end{equation}
are employed \cite{Houa}. Under these assumptions, wave solutions are obtained, and their physical properties, such as polarization and speed of waves, are discussed. In this paper, however, we want to relax one of these conditions and employ
\begin{equation}\label{constraints}
	K(0)\neq 0,\quad  K_{,\phi}(0)=0.
\end{equation} 
This will lead to a background curvature term originating from the minimum value of the scalar field potential, behaving like an effective cosmological constant. Hence, relaxing this condition effectively makes the spacetime asymptotically non flat. Our aim is to investigate how this choice changes the background geometry and affects the gravitational waves.

The variation of the action (\ref{Hornaction}) with respect to $g_{\mu\nu}$ and $\phi$ gives the field equations presented explicitly in \cite{Kobayashi}. In linearized approximation, in those field equations, we keep terms first order in $h_{\mu\nu}$ and $\varphi$ in \cite{Kobayashi}, and also use (\ref{Kexpansion}) to obtain
\begin{align}\label{leq1}
	&-\frac{1}{2} K(0)\eta_{\mu\nu} + G_4(0)G^{(1)}_{\mu\nu}(0)-G_{4,\phi}(0)(\partial_\mu\partial_\nu\varphi-\eta_{\mu\nu}\Box\varphi) = 0, \\
	&
	[K_{,X}(0)-2G_{3,\phi}(0)]\Box\varphi + K_{,\phi\phi}(0)\,\varphi + G_{4,\phi}(0)R^{(1)} = 0, \label{leq2}
\end{align}
where $G^{(1)}_{\mu\nu}$ is the linearized Einstein tensor and  $R^{(1)}$ is the linearized Ricci scalar. Here we use the notation $K_{,X}=\frac{\partial K}{\partial X}$ and $K_{,\phi}=\frac{\partial K}{\partial \phi}$ for a function K of  $\phi$ and $X$. Note also that all differential operators in the expressions now correspond to their flat Minkowski counterparts.

Using the  conditions (\ref{constraints}), the scalar field equation  (\ref{leq2}) becomes
\begin{align}\label{massc1}
	(\Box-m_s^2)\varphi&= 2K(0)\chi, 
\end{align}
with
\begin{equation}
	\chi=\frac{G_{4,\phi}(0)}{G_4(0)[K_{,X}(0)-2G_{3,\phi}(0)+3G^2_{4,\phi}(0)/G_4(0)]}.
\end{equation}
It is clear that the scalar field in (\ref{massc1}) is massive, where the mass of the scalar field  term, $m_s$, is defined  as,
\begin{equation}\label{ms}
	m_s^2=- \frac{ K_{,\phi\phi}(0)}{K_{,X}(0)-2G_{3,\phi}(0)+\frac{3G^2_{4,\phi}(0)}{G_4(0)}},
\end{equation}
which frequently occurs in 	Yukawa-type corrections to the Newtonian gravitational potential and mass parameter in the Klein-Gordon equation. For BD theory \cite{Brans1} ($K=0$) and Brans-Dicke Theory with a cosmological constant (BD$\Lambda$) \cite{Lorenz-Petzold,Endo,Kim} ($K\approx -2\Lambda_1 \phi)$, the scalar field is massless, i.e.  $m_s=0$ \cite{ Brans1,Ozer1} and for massive BD theory \cite{Alsing} $m_s^2=\frac{\phi_0}{3+2\omega}M''(\phi_0)$ with $M$ being the potential of this theory and for $f(R)$ theory, $m_s^2=\frac{f'(0)}{3f''(0)}$ \cite{Sotiriou}. In general, our analysis assumes a massive scalar field. The massless case can be recovered by setting  $m_s=0$, which corresponds to a vanishing, constant, or at most linear scalar potential in \(\phi\).

To simplify the field equation (\ref{leq1}), we introduce an auxiliary tensor field, whose appropriate versions are frequently used in scalar tensor \cite{Alsing} and Horndeski  theories \cite{Houa}, as
\begin{equation}\label{hmunudef}
	h_{\mu\nu}= \theta_{\mu\nu} - \frac{1}{2}\eta_{\mu\nu}\theta-\eta_{\mu\nu}\sigma\varphi,
\end{equation}
where $\sigma=\frac{G_4,\phi(0)}{G_4(0)}$. Using this notation, the field equations (\ref{leq1}) take the more compact form
\begin{equation}
	\partial_\rho\partial_{(\mu}\theta^\rho_{\nu)} - \frac{1}{2}\Box\theta_{\mu\nu} - \frac{1}{2}\eta_{\mu\nu}\partial_\rho\partial_\sigma\theta^{\rho\sigma}= 0.
\end{equation}
Here, the symmetric and second rank tensor field $\theta_{\mu\nu}$ makes the tensorial part of the equations easier to solve. Moreover, if we introduce a Lorenz gauge 
\begin{equation}\label{gauge}
	\partial^\mu \theta_{\mu\nu}=0,
\end{equation}
then, the field equations of linearized Horndeski theory reduce to the scalar equation given by  (\ref{massc1}), and the tensorial equation given as: 
\begin{align}\label{Boxtheta}
	\Box\theta_{\mu\nu}&= - \frac{K(0)}{G_4(0)}\eta_{\mu\nu},
\end{align}
where we have an inhomogeneous D'Alembert equation where $K(0)$ acts as a source term. 
As can be seen from equation (\ref{Boxtheta}), the tensor equation has a similar structure to General Relativity theory with a cosmological constant (GR$\Lambda$) \cite{Bernabeu} and Brans–Dicke theory with a cosmological constant (BD$\Lambda$) \cite{Ozer1}, and according to equation (\ref{massc1}), the scalar field is, in general,  both massive and has a source term. When we compare the equations (\ref{massc1},\ref{Boxtheta}) with corresponding  asymptotically flat space-time equations with $K(0)=0$, there are terms related to $K(0)$, $G_{4}(0)$ and  $G_{4,\phi}(0)$ terms in the equations. In other words, the field equations depend linearly on the scalar field nonminimally coupled to gravity and the potential that creates the background curvature. Moreover, by comparing corresponding equations of \cite{Bernabeu} in Lorenz gauge,  we see that $K(0)$ must be negative and   $K(0)\rightarrow -2\Lambda\, G_4(0) $ to have a de Sitter-like theory in the GR$\Lambda$ limit.
Having separated the scalar and tensorial parts helps us to obtain solutions more easily. However, the effects of the scalar field will emerge after obtaining the metric perturbation tensor using (\ref{hmunudef}), which clearly involves both tensorial and scalar parts in the physical spacetime metric.
Having obtained the necessary field equations,  let us now obtain their solutions in the next section.

\section{BACKGROUND AND GRAVITATIONAL WAVE SOLUTIONS ON A CURVED SPACETIME IN HORNDESKI THEORY WITH A POTENTIAL}

In this section, we aim to obtain gravitational waves in Horndeski theory in the weak field regime in the presence of a scalar potential term. Since we will not study strong-field effects, it is convenient to use the linearized version of the field equations. This choice will help us to investigate generic features of Horndeski theory in this regime without specifying exact forms of the scalar functions $G_i(\phi)$ and potential term $K(\phi)$. To apply linearized field equations, we need to  determine the background geometry, which clearly should be a solution of these equations.  In GR, we know that the vacuum solution with a positive cosmological constant is the de Sitter spacetime. In the presence of a non-vanishing minimum of the potential, $K(0)$, we expect that the corresponding background solution could be a linearized version of the de Sitter spacetime. Since it is not known whether de Sitter spacetime is a solution of the full Horndeski theory for arbitrary values of the functions $K,G_i$ in the Horndeski Lagrangian given in Eq. (\ref{Hornaction}), we instead derive the background solution within the linearized Horndeski theory and then examine whether it corresponds to a de Sitter–like spacetime. Although the background solution is {not asymptotically flat}, in the regime $H L \ll 1$ it can be locally expressed as Minkowski spacetime plus perturbations. However, our analysis focuses on the local weak-field regime, characterized by  $H L \ll 1$, where $H \approx\sqrt{|K(0)|}$ is the effective Hubble parameter and $L$ denotes the characteristic size of the region in which the gravitational wave is generated, propagates, and is detected. This scale is set by the typical length scales of the system, namely the gravitational-wave wavelength, the size and distance of the source, and the spatial scale of the detector. In practice, we take 
$L$ to correspond to the largest among these, which ensures that the entire physical region remains in the weak-field regime. In this limit, the curvature effects of the  background are extremely small, and the metric can be represented locally as flat Minkowski spacetime plus corrections of order $O(H^2)$. Therefore, it is  a consistent and practical choice to adopt the Minkowski spacetime as the zeroth-order background, and to include the effects of the minimum of the potential, $K(0)$, as an effective perturbative source term in the linearized field equations. In other words, the background geometry due to the existence of the minimum of the potential will be considered as a source term generating the metric perturbation tensor $h_{\mu\nu}^B$ and scalar perturbation function $\varphi^B$.  This method corresponds to local flat-space approximation of the background curvature induced by $K(0)$. Similar approach has been widely used in previous studies to investigate the impact of a cosmological constant on gravitational waves and local spacetime geometry in GR as well as in BD theories \cite{Bernabeu1,Arraut,Espriu,NafJetzer,Ozer1}. After determining the background geometry and the scalar field in this way, we can then obtain linearized GW solutions in this  background, thanks to the linearity of the field equations. Now our total metric becomes $g_{\mu\nu}=g_{\mu\nu}^B=\eta_{\mu\nu}+h_{\mu\nu}^B$,  compatible with Horndeski theory with a potential. Since we want to stay in the linearized level, we ignore second-order terms such as $K(0) h_{\mu\nu}$ in this work. This means that, our background metric is now $g_{\mu\nu}^B$, and we keep the terms linear by ignoring terms like $K(0)h_{\mu\nu}^{B,W} $, $K(0) \varphi^{B,W}$, where $h_{\mu\nu}^W$ and $\varphi^W$ represent perturbation terms related to gravitational waves, which are also assumed to be small. This does not assume a flat background, it is simply the standard perturbative method. As a consequence, the background–wave interaction does not modify the local form of the wave equation, but it manifests itself through observable quantities computed from the full metric. Since the total metric is not flat, $h_{\mu\nu}^B$ induces genuine curvature. Consequently, observable effects such as geodesic deviation must be computed in this curved geometry rather than in Minkowski spacetime. In summary, the wave equation looks the same as in flat space at linear order, but the background curvature controls the  GWs propagation and other observable effects such as frequency, wavelength and amplitude of the waves.

Once we have obtained these background and wave solutions in the coordinates compatible with these field equations (\ref{massc1},\ref{Boxtheta}) and chosen gauge (\ref{gauge}), we can then transform them to the coordinates where we make our observations, such as comoving cosmological coordinates, as we will discuss in section. In these coordinates, the first-order interaction between the minimum of the potential and GWs is clearly seen. This approach has been successfully applied in previous works to study the effects of the cosmological constant on GWs \cite{Bernabeu1,Arraut,Espriu,NafJetzer} and on the local spacetime geometry \cite{Bernabeu,Chauvineau} within GR and it has been recently extended to the BD theory \cite{Ozer,Ozer1}.     
As we have discussed above, here, we will obtain wave solutions of the Horndeski theory in the presence of background curvature due to the minimum of the potential, $K(0)$, and investigate the contributions of the background curvature (mimicking a non-zero cosmological constant or potential) to these field equations and their solutions. 
Since the field equations are linear, we can now decompose the spacetime metric into the background perturbation and the gravitational-wave perturbation generated by the background curvature.
\begin{align}
	g_{\mu\nu}= \eta_{\mu\nu} +{h^B} _{\mu\nu}+{h^W}_{\mu\nu},
\end{align}
where
\begin{equation}
	|h^{B,W}|\ll 1.
\end{equation}
Here ${h^B} _{\mu\nu}$ is the background perturbation and ${h^W}_{\mu\nu}$ is the gravitational wave perturbation. We can also expand the scalar field and $\theta_{\mu\nu}$ tensor in a similar way \cite{Ozer1}
\begin{equation}
	\theta_{\mu\nu}={\theta^B}_{\mu\nu}+{\theta^W}_{\mu\nu},
\end{equation}
Similarly, the scalar field is decomposed as 
\begin{equation}
	\varphi=\varphi^B+\varphi^W,
\end{equation}
with $\varphi^{B,W}\ll 1$.
Both background modification and gravitational wave perturbation affect the field equations. Hence, below we present the effects of both background modification and gravitational wave perturbation on the linearized field equations and solutions, respectively.

\subsection{Background Solutions}

In this subsection, we will be concerned with background linearized solutions by ignoring possible ripples in the spacetime.
In order to find the linearized background metric and scalar field due to $K(0)$,  we use the method used in GR \cite{Bernabeu,Bernabeu1} and BD \cite{Ozer} theories.
For background solutions generated by the minimum of the potential $K(0)$, linearized tensor and scalar equations  become,
\begin{equation}\label{fetbg}
	\Box\theta^B_{\mu\nu}=-\frac{K(0)}{G_4(0)} \eta_{\mu\nu},
\end{equation}
\begin{equation}\label{fescbg}
	(\Box-m_s^2)\varphi^B= 2K(0)\chi,
\end{equation}
together with the gauge condition
\begin{equation}\label{gaugebg}
	{\theta^{B\, \mu \nu}}_{,\mu}=0.
\end{equation}
Here we assume the solutions of (\ref{fetbg}) are quadratic in coordinates, as in \cite{Bernabeu}. Together with the Lorenz gauge given by Eq.(\ref{gaugebg}), the solutions of the field equations (\ref{fetbg}) take the following form,
\begin{equation}\label{thetaB}
	\theta^B_{\mu\nu}=\frac{K(0)}{18G_4(0)}x_\mu x_\nu-\frac{5K(0)}{36G_4(0)}\eta_{\mu\nu}x^2,
\end{equation}
where $x^2=\sum_\mu{x^\mu}^2$.

	For the solutions of scalar field equation, (\ref{fescbg}), we have to distinguish massless and massive cases. For both cases, however, we will use the following solution: 
	\begin{equation} \label{varphiB}
	\varphi^B=\frac{K(0)\,\chi\,  r^2}{3},
	\end{equation}
	which is also quadratic in coordinates. For massless case, $m_s=0$,   Eq. (\ref{varphiB}) is an exact solution.  If the scalar field has nonvanishing mass term, $m_s\neq 0$, then  Eq. (\ref{varphiB})  is an approximate solution valid througout the linearized approximation we use in this work. To see this let us investigate the full solution of (\ref{fescbg}). The general form of the solution  of this equation reads
	$
	\varphi^B_{(m_s\neq 0)}=-2K(0)\chi/m_s^2+ c_1 \exp[-m_s r]/r+c_2  \exp[m_s r]/(2 m_s r),
	$
	where $c_1$ and $c_2$ are arbitrary integration constants. This solution diverge when $m_s\rightarrow 0$. We can fix the integration constants $c_1,c_2$  by demanding both the regularity at $r\rightarrow 0$ and a smooth transition to $m_s\rightarrow 0$ as  boundary conditions. Expanding the expression in series and demanding such behaviors, we find that $c_1=-K(0)\chi/m_s^3$, $c_2=2K(0) \chi/m_s^2$, leading the solution to  $\varphi^B_{(m_s\neq 0)}= -2 K(0)\chi/{m_s^2}+2 K(0)\chi \sinh(m_s r)/(m_s^3 r)$. Series expanding this solution up to order $\mathcal O[K(0)]$ and neglecting the terms of the order  $\mathcal{O}[m_s^2 K(0)]$ and higher,  we find that for \emph{both massless and massive} Horndeski theories, the solution of the scalar perturbation equation (\ref{fescbg})  becomes as Eq. (\ref{varphiB})
	up to the  linearized order that we are considering.  In other words, in the weak field regime, the curvature associated with the background is governed primarily by the constant part of the potential, and the term proportional to $m_s^2 K(0)$ remains subdominant. Therefore, at leading order we can neglect this term and obtain the simplified background equation given by (\ref{varphiB}). Indeed, putting back the solution (\ref{varphiB}) the use of Eq. (\ref{varphiB}) as a background scalar field solution of linearized Horndeski theory is justified. 

 Putting $\theta_{\mu\nu}^B$  back into the metric perturbation tensor via (\ref{hmunudef}), we have obtained the full background metric perturbation as
\begin{equation}\label{hbackground}
	{h^B}_{\mu\nu}=\frac{K(0)}{18G_4(0)}x_\mu x_\nu+\frac{K(0)}{9G_4(0)}\eta_{\mu\nu} x^2-\eta_{\mu\nu}\,\sigma\,\varphi^B.
\end{equation}

Since we investigate these solutions under linearized approximation, we have chosenthe scalar field as independent of time as in the Newtonian form. The tensor part can be selected depending on time, as it can be converted to a static form under certain coordinate transformations. The solution given by Eq.  (\ref{hbackground}) is formally reduced to the corresponding one obtained in the GR$\Lambda$ and BD$\Lambda$ theories \cite{Bernabeu,Ozer1} under appropriate limits. Using Eq. (\ref{hbackground}), we find the space-time metric as,
\begin{equation}\label{backmetr}
	\begin{split}
		ds^2=-&\left[1-\frac{K(0)}{18G_4(0)}\left(3t^2-2r^2 \right)-\sigma\varphi^B\right]dt^2 + \sum_{i=1}^3 \left\{1+\frac{K(0)}{18G_4(0)}\left[-2t^2+2r^2+\left(x^{i}\right)^2\right]-\sigma\varphi^B\right\}(dx^{i})^2\\
		&-\sum_{i=1}^3\frac{K(0)}{9G_4(0)}t\, x^i\, dt dx^i + \sum_{\substack{i,j=1,\\ i\neq j}}^3  \frac{K(0)}{9G_4(0)}x^i x^j dx^i dx^j.  
	\end{split} 
\end{equation}

 This line element is neither homogeneous nor isotropic. In order to compare this metric with observable results, we need to bring it to a homogeneous and isotropic form. In the following, we will 	use several coordinate transformations to bring the solution into isotropic or Schwarzschild type coordinates. To do this we will adapt the transformations presented in \cite{Bernabeu1, Bernabeu} and ignore  all the terms higher than first order in  $K(0)$ in the forthcoming expressions, in accordance with the weak field approximation we are using in this paper. First, let's convert the metric into a static form with the following coordinate transformations,
\begin{eqnarray}
	&&	x=\tilde{x}-\frac{K(0)}{18G_4(0)}\Big(-\tilde{t}^2-\frac{\tilde{x}^2}{2}+\frac{(\tilde{y}^2+\tilde{z}^2)}{4}\Big)\tilde{x}, \nonumber\\
	&&	y=\tilde{y}-\frac{K(0)}{18G_4(0)}\Big(-\tilde{t}^2-\frac{\tilde{y}^2}{2}+\frac{(\tilde{x}^2+\tilde{z}^2)}{4}\Big)\tilde{y},\\
	&&	z=\tilde{z}-\frac{K(0)}{18G_4(0)}\Big(-\tilde{t}^2-\frac{\tilde{z}^2}{2}+\frac{(\tilde{x}^2+\tilde{y}^2)}{4}\Big)\tilde{z},\nonumber \\
	&&	t=\tilde{t}+\frac{K(0)}{36G_4(0)}(\tilde{t}^2+\tilde{r}^2\tilde{t})\,\tilde{t}\nonumber
\end{eqnarray}
where $\tilde{r}^2=\tilde{x}^2+\tilde{y}^2+\tilde{z}^2$. Under these coordinate transformations, the resulting metric and scalar field solution become,
\begin{equation}
	ds^2=-\Big[1+\frac{K(0)}{6G_4(0)}\tilde{r}^2-\sigma\tilde{\varphi}^B\Big]\tilde{dt^2}+
	\sum_{i=1}^3
	\left[1+\frac{K(0)}{12G_4(0)}(\tilde{r}^2+3\tilde{x}_i^2)-\sigma\tilde{\varphi}^B\right]d\tilde{x_i}^2,
\end{equation}
\begin{equation}
	\tilde{\varphi}^B=\frac{K(0)\chi }{3}\tilde{r}^2.
\end{equation}
Now the metric tensor is made diagonal, but not static yet.
With the following coordinate transformations, 
\begin{eqnarray}	
	\tilde{x}=x'-\frac{K(0)}{24G_4(0)}x'^3, \nonumber\\
	\tilde{y}=y'-\frac{K(0)}{24G_4(0)}y'^3, \\
	\tilde{z}=z'-\frac{K(0)}{24G_4(0)}z'^3 ,\nonumber
\end{eqnarray}
\begin{equation}	
	\tilde{t}=t'\nonumber,
\end{equation}
and some algebraic manipulations lead to a static and isotropic form for the line element
\begin{equation}\label{metrbackisot}
	ds^2=-\left[1+\frac{K(0)}{6G_4(0)}r'^2-\sigma\varphi'^B\right]dt'^2+ \left[ 1+\frac{K(0)}{12G_4(0)}r'^2-\sigma\varphi'^B \right]\left(r'^2+r'^2 d\Omega^2\right),
\end{equation}
with $ d\Omega^2$ being the metric of the unit sphere. Furthermore, we  will perform another coordinate transformation to obtain the Schwarzschild de Sitter-like metric modified by the scalar field to the first order of $h_{\mu\nu}$ and $K(0)$,
\begin{eqnarray}
	r'=\bar{r}\left(1-\frac{K(0)}{24\,G_4(0)}\bar r^2+\frac{\sigma\varphi'^B}{2}\right),
\end{eqnarray}
\begin{equation}
	t'=\bar t. \nonumber
\end{equation}
where $\bar{r}$ is the usual Schwarzschild radial coordinate where in these coordinate the area of a sphere with constant radius is $4 \pi \bar r^2$, and  $\bar t$ refers to the Schwarzschild time coordinate.
This transformation brings the line element (\ref{metrbackisot}) to the following expression,
\begin{equation}\label{metrbacksch}
	ds^2=-\left[1+\frac{K(0)}{6G_4(0)}\bar r^2-\sigma \varphi^B\right] d\bar t^2+\left[1-\frac{K(0)}{6G_4(0)}\bar r^2+\sigma \alpha\, \bar r\right]d\bar r^2+ \bar r^2 d\Omega^2,
\end{equation}
where we have defined
\begin{equation}
	\alpha=\frac{d\varphi^B}{d\bar r},
\end{equation}
and the scalar field becomes,
\begin{equation}
	\varphi^B=\frac{K(0)\chi }{3} \bar r^2.
\end{equation}
This solution reduces to the linearized dS metric in Schwarzschild coordinates, when the scalar perturbation vanishes. So we have eliminated the time dependence of the metric and brought it into various static forms (\ref{metrbackisot},\ref{metrbacksch}) for future applications. It should be emphasized that the simple metric forms given in Eqs. (\ref{metrbackisot}) and (\ref{metrbacksch}) do not arise from a new physical approximation, but rather follow from the weak-field approximation adopted in this work. In this framework, the metric and the Horndeski scalar field are expanded around their Minkowski values and only terms up to first order in the perturbations and in the minimum of the potential term are retained. The coordinate transformations employed below are implemented perturbatively at the same order and reflect the gauge freedom of the linearized theory. Consequently, the resulting metrics represent different local, lowest-order representations of the same linear solution of the background geometry.

\subsection{Wave Solutions}

We now focus on the oscillatory wave part  of the field equations. We will find wave-like solutions by taking into account the GW perturbation that may occur in the presence of a source that causes fluctuations in the space-time geometry. In this case, the gauge condition can be written,
\begin{equation}\label{gaugewave}
	{\theta^{W \mu \nu}}_{,\mu}=0.
\end{equation}
The homogenous part of the total field equation (\ref{Boxtheta}) represents the wave part of the equation
\begin{equation}\label{homtenwave}
	\Box{\theta^W}_{\mu\nu}=0.
\end{equation}
One sees from Eq. (\ref{homtenwave}) that the gravitational wave type solutions of the linearized field equations in the Lorenz gauge are identical to those existing in flat space-time. The same decomposition holds for the scalar field, so the homogeneous part of the total scalar field equation (\ref{massc1}) gives  a massive scalar  wave equation
\begin{equation}\label{homscwave}
	(\Box-m_s^2)\varphi^W=0.
\end{equation}
The mass term we have ignored in the linearized  background solution cannot be ignored in the wave equation and  it modifies the dynamics of the propagating scalar mode. This is because  $\varphi^W$ describes rapidly varying fluctuations, making $m_s^2\varphi^W$  comparable to  $\Box\varphi^W$ and therefore non-negligible. Consequently, the scalar-field mass plays no role at the background level and emerges solely through the wave perturbation, as we will discuss below.

The solutions of the tensor wave equations (\ref{homtenwave})  admit the plane-wave solutions,
\begin{equation}\label{wavetensoln}
	{\theta^W}_{\mu \nu}=A_{\mu\nu}\sin kx+B_{\mu\nu}\cos kx,
\end{equation}
where $ A_{\mu\nu}$ and $B_{\mu\nu}$ are amplitude tensors, the four-vector   $k=(\nu,k_x,k_y,k_z)$  is the wave vector whose timelike component gives the frequency of the wave, and we use the shorthand notation $kx=k_\mu x^\mu$. Plugging the solution (\ref{wavetensoln})  into the Eq. (\ref{homtenwave}), we obtain the condition,
\begin{equation}
	k_\mu k^\mu=0.
\end{equation}
This shows the fact that the tensorial waves propagate at the speed of light. Using the Lorenz gauge condition (\ref{gaugewave}) we see that  
\begin{equation}
	k_\mu {A^W}^\mu_{\nu}=	k_\mu {B^W}^\mu_ {\nu}=0,
\end{equation}
implying that the amplitude tensors are orthogonal (transverse) to the direction of the propagation of the waves. Moreover, It is possible use of the remaining gauge freedom to impose a transverse-traceless (TT) gauge such that
\begin{equation}
	{A^W}^\mu_ {\ \mu}= {B^W}^\mu_ {\ \mu}=0,\quad A_{0i}=B_{0i}=0.
\end{equation}

The solution of the  scalar field wave equation (\ref{homscwave}) can be written as
\begin{equation}
	\varphi^W=\varphi_0\, e^{iqx},
\end{equation}
with the wave vector  $q$ is designated to be different from than the tensorial four-wave vector given by
\begin{equation}
	q^\mu=(\omega ,\tilde k_x,\tilde k_y,\tilde k_z).
\end{equation}
 The frequency $\omega$ and wave vector components $\tilde k_i$ of the scalar wave, obey the dispersion relation
\begin{equation}\label{dispersionsc}
	-q_\mu\, q^\mu=\omega^2-(\tilde k_x^2+\tilde k_y^2+\tilde k_z^2)=\omega^2-\tilde k^2=m_s^2.
\end{equation}
Here and in the following expressions, we will consider the real part of $\varphi^W$. Eq. (\ref{dispersionsc}) implies that the dispersion law of light is not satisfied in the massive linearized Horndeski theory. The effect of the potential that produces the effective mass in the wave solutions comes from the solution of the scalar field equation. The most important result of this solution is that scalar waves move more slowly than tensor waves in this theory. So, in a way, the mass of the scalar field acts as a slowing medium for scalar waves.
The total metric perturbation corresponding to gravitational waves, in which both scalar and tensorial solutions contribute, can be written as,
\begin{equation}\label{hwave}
	{h^W}_{\mu\nu}=A_{\mu\nu}\sin kx+B_{\mu\nu} \cos kx-\eta^{\mu\nu}\sigma\varphi_0\, e^{iqx}.
\end{equation}
The first two terms in this expression represent the tensor waves, and the last term represents the scalar wave. 

Now we can combine background (\ref{hbackground}) and wave (\ref{hwave}) solutions into the spacetime metric, since the total solution of the linearized field equation is the sum of the background solution and the wave solution. 
Hence, we have the total metric perturbation solution can be written as
\begin{equation}\label{htotal}
	{h}_{\mu\nu}= \frac{K(0)}{18G_4(0)}x_\mu x_\nu+\frac{K(0)}{9G_4(0)}\eta_{\mu\nu}x^2-\frac{\eta_{\mu\nu}\sigma K(0)\chi r^2}{3}+A_{\mu\nu}\sin kx+B_{\mu\nu}\cos kx-\eta^{\mu\nu}\sigma\varphi_0e^{iqx},
\end{equation}
and the total scalar field solution is given by
\begin{equation}\label{sctotal}
	\varphi=\frac{
		K(0)\chi r^2}{3}+\varphi_0e^{i\,q \,x}.
\end{equation}
As a result, we have obtained the solutions of the gravitational wave equations in the linear order in the Horndeski theory {\it under the presence of background curvature} due to a nontrivial potential term in the action. The result in (\ref{htotal}) reduces to corresponding solutions in GR \cite{Bernabeu1} and BD$\Lambda$ \cite{Ozer} theories. When the background curvature term $K(0)$ vanishes, the solution reduces to linearized Horndeski gravitational waves \cite{Houa}. Unlike GR, and similar to BD theory, the solutions are accompanied by a scalar wave as well as tensor waves. In this theory, since the potential of the scalar field adds mass to the scalar field, the scalar waves move slower than the speed of light. Tensor waves, on the other hand, are compatible with the predictions of the GR theory, and the waves move at the speed of light. The total effect of the GWs in Horndeski theory is a combination of both scalar and tensorial waves, as we will discuss in the next section regarding the effects of the background curvature generated from the minimum of the potential.

Note that although  the wave part in (\ref{htotal}) resembles the flat space counterparts, when these waves propagate, they will be affected by the curvature originated from  $K(0)$, the minimum of the potential. Therefore,  the waves  will interact with background curvature via the full spacetime metric,  which also includes contributions of Horndeski scalar field due to $K(0)$. In the following sections we will investigate how the minimum of the potential affects the gravitational waves nontrivially, even in the linearized level. In other words, we will determine how the minimum of the potential make these waves different then waves in Minkowski spacetime.

\section{The Effects of GWs on the Free Particles in Massive Horndeski Theory}

In this section, we will investigate how gravitational waves interact with test particles or detectors in the space-time geometry in the  linearized Horndeski theory in the presence of a constant background curvature originated from the minimum of the potential and  then analyze the polarization properties of these waves. Since the polarization properties of GWs of Horndeski theory are well known \cite{Houa} in a flat background, we aim to focus mainly on the nonflat background due to the potential of the scalar field. The test particles follow the geodesics of that spacetime in the curved spacetime. Therefore, we will use geodesic deviation equations to study the effects of GWs on a free particle. Due to the equivalence principle, no gravitational field acts on a particle located at a single point in space-time. When a GW passes through a point, it does not change the coordinates of that point, but during passage, it changes the physical distance between two distinct points located in the spacetime. Therefore, a single particle is not affected by space-time fluctuations. The easiest way to understand the physical effects of GWs on matter is to consider the relative motion of two neighboring test particles, both initially at rest, in free fall \cite{Schutz}. We know that the wave vector $k$ is a null vector, so for a wave traveling along the $z$ direction with speed of light in vacuum, the wave vector corresponding to tensorial waves can be written as
\begin{equation}
	k^\mu=(\nu,0,0,\nu), 
\end{equation}
where $\nu$  is the frequency of the tensorial waves. Similarly, the wave vector of the scalar waves, obeying the dispersion relation (\ref{dispersionsc}) giving $\omega^2-\tilde k^2=m_s^2$, takes the form
\begin{equation}
	q^\mu=(\omega,0,0,\tilde k)=(\omega,0,0,\sqrt{\omega^2-m_s^2}), 
\end{equation}
for frequency $\omega$ of the scalar waves. The propagation speed of scalar waves is less than speed of light $v=\sqrt{\omega^2-m_s^2}/\omega$ \cite{Houa}. For calculational simplicity, and without loss of generality, the general solution of the total field equation (\ref{htotal}) for a wave propagating in the $+z$ direction can be written as
\begin{equation}\label{htotal1}
	h_{\mu\nu}=h_{\mu\nu}^B+h_{\mu\nu}^W=\frac{K(0)}{18G_4(0)}x_\mu x_\nu+\frac{K(0)}{9G_4(0)}\eta_{\mu\nu}x^2-\frac{\eta_{\mu\nu}\sigma K(0)\chi r^2 }{3}+\tilde A_{\mu\nu}\cos \nu(t-z)-\eta_{\mu\nu}\,\sigma\,\varphi_0\, e^{i(\tilde k z-\omega t)},
\end{equation}
by employing relevant dispersion relations, and combining $A_{\mu\nu}$ and $B_{\mu\nu}$ into $\tilde A_{\mu\nu}$ by using properties of trigonometric functions and taking emerged phase constants vanishing, which does not have any effect on the present discussion.

Let us calculate the geodesic deviation equations caused by the GW propagating in the $+z$ direction. The geodesic deviation equation, for example given by \cite{Rindler}, is
\begin{equation}
	\frac{d^2\zeta^\alpha}{d\tau^2}=R^\alpha_{\beta\vartheta\lambda} U^\beta U^\vartheta \zeta^\lambda.
\end{equation}
Here $\zeta^\alpha$ is the vector connecting the particles, $U^\mu$ is the four velocity of the particles, and in the rest frame of the particle it can be written as $U^\mu=(1,0,0,0)$. The perturbation tensor (\ref{htotal1}) implies that as the wave propagates, it produces an oscillating Riemann tensor.

Using the linearized gauge-invariant definition of Riemann tensor and Eq. (\ref{htotal1}), and adopting an appropriate gauge choice for the tensor sector, we can show that the only nonvanishing components of the Riemann tensor in TT gauge are,
\begin{eqnarray}\label{Riemcomp}
	&&R^x_{0x0}=\frac{1}{2}\left(\nu^2 \bar A_{xx}-\omega^2\sigma \bar{\varphi}- V\right),\nonumber \\
	&&	R^y_{0y0}=\frac{1}{2}\left (- \nu^2 \bar A_{xx}-\omega^2 \sigma \bar{\varphi}- V \right),\nonumber \\
	&&		R^z_{0z0}=\frac{1}{2}\left(-m_s^2 	\sigma \bar{\varphi}- V \right), \label{Riemcomp} \\
	&&	R^x_{0y0}=\frac{1}{2}\nu^2\bar{A_{xy}},\quad 		R^z_{0x0}=R^z_{0y0}=0,\nonumber
\end{eqnarray}
agreeing with \cite{Houa} for $\Lambda=0$, with the abbreviations
\begin{equation}
	\bar A_{\mu\nu}=\tilde A_{\mu\nu} \cos \nu\,(t-z), \quad \bar\varphi =\varphi_0\, e^{iqx}, \quad V=- \frac{K(0)}{3G_4(0)}\left(1-2\sigma\chi G_4(0)\right).
\end{equation}
In these expressions, we take into account the real part of the term  $e^{iqx}$. Let us assume that the first particle is at the origin and second one is at the point  $\zeta^i=(0,\zeta,0,0,)$ and that the particles are initially at rest.
In this case, the relative accelerations of the freely falling test particles are found to be
\begin{eqnarray}
	&&	\frac{d^2\zeta^x}{d\tau^2}=-\frac{\zeta}{2}\left(\nu^2 \bar A_{xx}-\omega^2\sigma \bar{\varphi}-V\right),\nonumber \\
	&&	\frac{d^2\zeta^y}{d\tau^2}=-\frac{\zeta}{2}\nu^2\bar A_{xy},\\
	&&	\frac{d^2\zeta^z}{d\tau^2}=0.\nonumber
\end{eqnarray}
Similarly, two particles initially separated by $\zeta$ in the $y$ direction with similar settings obey the
equations
\begin{eqnarray}
	&&	\frac{d^2\zeta^x}{d\tau^2}=-\frac{\zeta}{2}\nu^2\bar A_{xy},\nonumber \\
	&&	\frac{d^2\zeta^y}{d\tau^2}=-\frac{\zeta}{2}\left (- \nu^2 \bar A_{xx}-\omega^2 \sigma \bar{\varphi}-V \right),\\
	&&	\frac{d^2\zeta^z}{d\tau^2}=0.\nonumber
\end{eqnarray}

Thirdly, the relative acceleration of two nearby freely falling particles seperated by $\zeta$ in the z direction becomes,
\begin{eqnarray}
	&&\frac{d^2\zeta^x}{d\tau^2}=0,\nonumber \\	
	&&\frac{d^2\zeta^y}{d\tau^2}=0,\\
	&&\frac{d^2\zeta^z}{d\tau^2}=-\frac{\zeta}{2}\left(-m_s^2 	\sigma \bar{\varphi}-V \right).\nonumber
\end{eqnarray}

The terms $\bar A_{xx}$ and $\bar A_{xy}$ in the geodesic deviation equations denotes the usual plus and cross polarizations of the gravitational wave, as in the GR theory. The  terms involving $\bar \varphi$ indicate the contribution of the scalar field. In our analysis, GW polarization states are identified only from the oscillatory, wave-dependent contributions to the Riemann tensor. The term proportional to $V=-\frac{K(0)}{3G_4(0)}(1-2\sigma\chi \varphi)$  originates from the background curvature associated with the minimum of the scalar potential and leads to an isotropic contribution to the geodesic deviation equation. This term does not correspond to a propagating GW polarization mode. As can be seen from the geodesic deviation equations, the terms related to the $K(0)$ term cause the same acceleration in all directions. The observed behavior arises from a homogeneous background curvature rather than from anisotropic tidal effects caused by GW polarizations. Also, the presence of the $\omega^2\sigma \bar \varphi$  and $m_s^2\sigma\bar \varphi$  terms  denotes the existence of transverse breathing scalar mode and longitudinal scalar modes, respectively. Therefore, in addition to the usual tensor modes of GR,  there exists a single additional scalar polarization mode. If the scalar field is massive, this mode induces both transverse breathing and longitudinal components.  However, in the massless case, the scalar mode gives rise only to a transverse breathing component, in agreement with the the original  BD and BD$\Lambda$ theories \cite{Brans1, Ozer1}.  Since  $\omega >m_s$, transverse displacements are always greater than longitudinal displacements.  Consequently, in the massive Horndeski theory, GWs  have three polarization states \cite{Houa}, two of which show plus and cross polarization states induced by the spin 2 field, and one scalar mode inducing both transverse and longitudinal components, as in the massive BD theory \cite{Alsing}. The background curvature resulting from the potential acts as a cosmological constant and causes homogeneous expansion between particles. Therefore the conditions $V>0$, $K(0)<0$ and $2\sigma\chi G_4(0)<1$   should be met. These conditions reduces to $\Lambda>0$ and $\frac{2}{2\omega_{BD}+3}<1$ conditions in BD$\Lambda$ theory  and are compatible with the literature \cite{Ozer1}.

The wave part of the metric perturbation terms can also be expressed as,
\begin{equation}
	h^W_{\mu\nu}(t,z)=A^+(t,z)e^+_{\mu\nu} +A^\times(t,z) e^\times_{\mu\nu}+\Phi(t,z)\eta_{\mu\nu},
\end{equation}
with
\begin{equation}
	e^+_{\mu\nu}=\begin{pmatrix}
		0&0&0&0\\
		0&1&0&0\\
		0&0&-1&0\\
		0&0&0&0
	\end{pmatrix},\qquad
	e^\times_{\mu\nu}=\begin{pmatrix}
		0&0&0&0\\
		0&0&1&0\\
		0&1&0&0\\
		0&0&0&0
	\end{pmatrix},\qquad
	\eta_{\mu\nu}=\begin{pmatrix}
		-1&0&0&0\\
		0&1&0&0\\
		0&0&1&0\\
		0&0&0&1
	\end{pmatrix}.
\end{equation}\\
Here $e^+_{\mu\nu}$ and $e^\times_{\mu\nu}$ denote the plus and cross polarizations of the gravitational waves, and $A^+=\bar A_{xx}$, $A^\times=\bar A_{xy}$  and $\Phi(t,z)=\sigma\bar\phi$ are tensorial and scalar amplitudes of the waves, respectively.

Another way to see all this is to express the (electrical) components of the Riemann tensor (\ref{Riemcomp}) in matrix form. In this case, we get the following expression:
\begin{equation}
	R^i_{0j0}=\frac{1}{2}\begin{pmatrix}
		\nu^2A^+-\omega^2 \Phi-V&\nu^2A^\times&0&\\
		\nu^2A^\times&-\nu^2A^+-\omega^2\Phi-V&0\\
		0&0&-m_s^2\Phi-V&\\
	\end{pmatrix}\qquad
\end{equation}
The term proportional to V in the Riemann tensor represents a homogeneous background curvature contribution and should not be interpreted as an additional GW polarization mode. Note that in the limit $V=0$ the results agree with \cite{Houa}. The properties we mentioned above are clearly understood from the electrical part of the Riemann tensor. Briefly, we have the + and $\times$ tensor polarizations, breathing scalar and transverse scalar modes and $V$ term associated with the minimum potential $K(0)$ derived from the background solution and corresponding to the homogeneous expansion.

\section{Cosmological and observational effects of background curvature on gravitational waves}

We have seen from geodesic deviation equations that the effect of background curvature originating from the minimum of the potential leads to a homogeneous and isotropic expansion of any two distinct points of spacetime, mimicking an effective cosmological constant. Hence, this effect is surely cosmological. Therefore, in order to better understand this effect, it is preferable to express the background and wave solutions in comoving FLRW coordinates, as is done \cite{Bernabeu1} in GR$\Lambda$ theory. This transformation is necessary since we make observations of the cosmos in these coordinates.  Therefore, we apply the following coordinate transformations, adopted from \cite{Bernabeu1}, as 
\begin{equation}
	x^i=e^{H\, T}\,X^i, \label{transfx}
\end{equation}
\begin{equation}
	t=\frac{1}{2}H\, R^2+T, \label{transft}
\end{equation}
where T is the comoving cosmic time, $X^i=(X,Y,Z)$ are comoving spatial coordinates and $R=\sqrt{X^2+Y^2+Z^2}$ is the corresponding comoving radial distance. The constant $H$ is defined here as
\begin{equation}
	H=\sqrt{\frac{|K(0)|}{6\, G_4(0)}}
\end{equation}
which reduces the de-Sitter Hubble parameter in the GR limit. Applying eqs. (\ref{transfx},\ref{transft}) to the background solution give the unperturbed FLRW background described by the line element given by
\begin{equation}
ds^2=-dT^2+e^{2HT}(dX^2+dY^2+dZ^2). \quad 
\end{equation}	
 The perturbative contribution will be contained entirely in $h_{\mu\nu}^W$ and will be discussed below.

After some algebraic calculations, we find the transformed wave-like solution to the order $H$, i.e., $\sqrt{|K(0)|}$, as
\begin{equation*}
	h^W_{\mu\nu}=\begin{bmatrix}
		0&0&0&0\\
		0&A_{xx}\left(1+2H\, T\right)&A_{xy}\left(1+2H\,T\right)&0&\\
		0&A_{xy}\left(1+2H\,T\right)&-A_{xx}\left(1+2HT\right)&0\\
		0&0&0&0&\\
	\end{bmatrix}
	\times
	\\
	\cos\left[\nu(T-Z)+\nu H \left(\frac{Z^2}{2}-TZ\right)+O(H^2)\right]
\end{equation*}
\begin{equation*}
	+	\begin{bmatrix}
		\sigma\varphi_0&0&0&0\\
		0&-\sigma\varphi_0\Big(1+2H T\Big)&0&0&\\
		0&0&-\sigma\varphi_0\Big(1+2H T\Big)&0\\
		0&0&0&-\sigma\varphi_0\Big(1+2H T\Big)&\\
	\end{bmatrix}
	\times\\
\end{equation*}
\begin{equation}\label{wavefrw}
	\cos\left[\omega \left(T-\sqrt{1-\frac{m_s^2}{\omega^2}}Z\right)+\omega H \left(\frac{Z^2}{2}-\sqrt{1-\frac{m_s^2}{\omega^2}}TZ\right)+O(H^2)\right]
\end{equation}
where $\sigma\varphi_0$ term is the real part of the scalar amplitude. The perturbative term $h^W_{\mu\nu}$ should be viewed as added to the FLRW background generated in (62)–(63). Thus, the geometry derived from the coordinate transformations supplies the unperturbed metric and the complete field configuration is obtained once the perturbation is added to it. As can be seen Eq. (\ref{wavefrw}), the tensor part of gravitational waves behave exactly the same as in GR$\Lambda$ \cite{Bernabeu1}  and BD$\Lambda$ \cite{Ozer} theories .
We observe similar behavior with GR case \cite{Bernabeu1}, where 
de Siter (dS) solution in FLRW coordinates is not a solution of linearized equations.  Hence, in GR, applying an appropriate  coordinate transformation  in \cite{Bernabeu1} brings the  GR$\Lambda$ wave solution in the order of $\sqrt{\Lambda}$ and not $\Lambda$,  similar to what we have obtained here in which corresponding  transformation brings the wave solution to the order of  $\sqrt{|K(0)|}$ but not $K(0)$. To obtain the solution of the order of $K(0)$,  we need to obtain next order of solutions of the order of $K(0)h_{\mu\nu}$ in all equations. Since second order solutions is much more complicated in Horndeski theory, we will postpone this to future works.

Using the maximum of the waves, the two polarization degrees of freedom, because of the tensor modes, will obey the same dispersion relation given by
\begin{equation}
	\nu(T-Z)+\nu H\Big(\frac{Z^2}{2}-TZ\Big)=n\,\pi,
\end{equation}
and we obtain
\begin{equation}
	Z_{max}(n,T)\simeq T-\frac{n\pi}{\nu}-H\Big(\frac{T^2}{2}-\frac{n^2\pi^2}{2\nu^2}\Big).
\end{equation}
This will give us a phase velocity $v_f=dZ_{max}/dT=1-H T+O(H^2)<1$ smaller than one. However, this does not mean tensorial waves moving at a speed smaller than the speed of light. To see this, as in \cite{Bernabeu1}, we can calculate proper velocity $v=dl/dT=d/dT\left[(1+2H T )dZ_{max}\right]\simeq 1$, where  $dl^2=(1+2H T )dZ^2$  is the ruler distance. 

However, massive scalar modes behave differently from tensor modes. The maximum of the scalar modes can be found as
\begin{equation}
	\omega T -\tilde k Z +  H  \left(\omega \frac{Z^2}{2}-\tilde k TZ\right)=n\pi,
\end{equation}
which gives
\begin{equation}
	\tilde k Z_{max}=\omega T- n\pi-\left[ \frac{m_s^2\, n\pi}{ \tilde k^2}T+\frac{\tilde k^2-m_s^2}{2\tilde k^2}T^2 - \frac{n^2\pi^2}{2\tilde k^2}\right]	H. 
\end{equation}
The phase velocity for scalar waves becomes
\begin{equation}
	v_f=\frac{\omega}{\tilde k}-H\left(\frac{m_s^2 n \pi}{\tilde k^2}+\frac{\tilde k^2-m_s^2}{\tilde k^2}T \right)+O(H^2)	
\end{equation}	
which is less than the phase velocity of waves in a medium due to cosmological expansion.
However, since the scalar wave moves slower than the speed of light, the speed of these waves depends on the observer. In the vanishing mass limit, the phase velocities agree for scalar and tensorial terms, and both have proper velocity equal to the speed of light in this limit. 

One outcome of the existence of a minimum of the potential is to reveal the cosmological effects of the background effective cosmological term $H$, responsible for the accelerated expansion of the universe in our framework. The effect of this expansion is to stretch the amplitude of the waves with time, with the effective time-dependent amplitudes such as  $\left(1+2H\, T \right)$. Simply, the waves undergo Hubble expansion when they move away from the source. It also affects both frequency and wave number of both tensorial and scalar waves, which we discuss in the following subsection. The transformations given by equations (\ref{transfx},\ref{transft}) cause changes in the structure of both the amplitude and harmonic part of the wave. We discuss these effects and their possible observational outcomes in the following.

\subsection{A possible way of determininig Hubble constant or distance of the source from effective frequency}

We know that the current GW detectors can only determine the frequency of the waves. Hence, we can determine the frequency of a wave from these observations. Using the results of these detections and what we have obtained in this paper, can we deduce some observational outcomes?  Let us speculate on this in the following discussion.

We can define effective frequencies and wave numbers for tensorial and scalar waves from (\ref{wavefrw}) by defining
\begin{eqnarray}
	&&\cos\left[\nu T(1-H Z)-\nu Z \left(1- \frac{H}{2} Z \right)  \right]=\cos (\nu_{eff} T-k_{eff} Z), \label{harmten}\\
	&& \cos\left[ \omega T \left(1-\frac{\tilde k}{\omega} H Z\right) -\tilde k  Z \left(1-\frac{\omega}{\tilde k} \frac{H}{2} Z \right) \right]=\cos(\omega_{eff}T-\tilde k_{eff} Z), \label{harmsc}
\end{eqnarray}
for the tensorial and scalar parts, respectively. The tensorial part agrees with the GR$\Lambda$ case  \cite{Bernabeu1} in the appropriate limit. Note that the effective frequency and wave number depend on the minimum of the potential and effective Newton constant $G_4(0)$. From the expression (\ref{harmten}),  we see that for Horndeski theory, the effective frequency and wave number  for tensorial waves are given  by
\begin{eqnarray}\label{efffreqtn}
	\nu_{eff}=\nu(1-HZ), \quad k_{eff}=\nu(1-HZ/2),
\end{eqnarray}
similar to  GR$\Lambda$ case \cite{Bernabeu1}. The effective frequency expression is compatible with the cosmological redshift the waves experienced during expansion of the universe, as $\nu_{eff}=\nu/(1+z)$ with $z=H Z$. However, the wave number does not follow such an expression.  They are different from each other, since $\nu_{eff}\neq k_{eff}$. Hence, contrary to expectation and as noticed before for GR \cite{Bernabeu1}, the accelerated expansion due to the effective cosmological constant leads to differences in time evolution and spatial elongation of tensorial waves differently. This difference is noticed first in GR \cite{Bernabeu1}. This difference may help us to determine the effective cosmological constant (or Hubble parameter) from gravitational wave observations. We can define effective frequency shift from (\ref{efffreqtn}) as
\begin{equation}
	\frac{\Delta \nu}{\nu}\equiv \frac{\nu-\nu_{eff}}{\nu}=H\,Z. \label{tenfreqshift}
\end{equation}	
Hence, knowing the frequency shift, the frequency of the waves at the source, $\nu$,  and the spatial distance of the source of gravitational waves, we can determine $H$. Alternatively, knowing the frequency shift and effective cosmological constant, we can determine the distance to the source of gravitational waves. Since $H$ is related to the minimum of the potential in Horndeski theory, we can determine it from gravitational wave observations. By appropriately choosing the potential and the relevant terms in Lagrangian (\ref{Hornactions}), such as $L_2$, this effective cosmological constant can be made equivalent to the actual cosmological constant in the GR limit. Hence, the above discussion can also help us to determine the cosmological constant and related Hubble parameter in the GR limit as well.

The corresponding effective frequency and wave number terms are more interesting for the massive scalar fields, obtained from (\ref{harmsc}) and given below:
\begin{eqnarray} \label{weffsc}
	&&\omega_{eff}=\omega \left(1-\frac{\tilde k}{\omega}H Z \right)=\omega \left(1-\sqrt{1-\frac{m_s^2 }{\omega^2}}H Z \right),\\
	&& k_{eff}=\tilde k   \left(1-\frac{\omega}{\tilde k} \frac{H Z}{2}  \right)=\tilde k   \left(1-\frac{1}{\sqrt{1-\frac{m_s^2 }{\omega^2}}} \frac{H Z}{2}  \right). \label{keffsc}
\end{eqnarray}
We see that when the scalar field is massless, the effective frequencies and wave number have the same relations with tensorial waves. However, for a massive scalar field,  the effect of the mass works oppositely for frequency and wave number. Namely, for a given frequency, increasing the mass increases the frequency redshift but decreases the redshift in the wave number. It is a very peculiar effect that has been noted for the first time, as far as we know.  

Similar to the previous discussion, we can define a frequency shift for scalar waves as
\begin{equation} \label{scfreqshift}
	\frac{\Delta \omega}{\omega}\equiv\frac{\omega-\omega_{eff}}{\omega}=\sqrt{1-\frac{m_s^2}{\omega^2}}H Z.
\end{equation}
Hence, determining the frequency shift for scalar waves, we can, for example, determine the mass of the scalar field, $m_s$, by knowing $H$ from observations on the tensorial part using $Z$, or vice versa, coming from the same source.

If the source is not a black hole-black hole (bh-bh) merger but contains a non-black element, such as a neutron star (ns), or a bh-ns or an ns-ns merger, we can determine some of the parameters from accompanying electromagnetic waves as well. This can help to further achieve more accurate results for these parameters.  For example, we suppose that we can determine $\nu$ and $\omega$ from the optical window, and $\nu_{eff}$ and $\omega_{eff}$ can be deduced from observations of gravitational detectors. Hence, using the presently accepted value of the effective cosmological constant and the coordinate distance of the source, which can also be obtained from the optical window, we can deduce the mass of the scalar field. Obtaining the effective cosmological constant $H$ can help us to limit the effective Newton constant $G_4(0)$ as well as the minimum of the potential  $K(0)$, which may help us to further restrict the Horndeski theory.  

\subsection{Observational results from effective wave number}

 As we have mentioned before, the present GW observatories in the LIGO-Virgo-Kagra collaboration can define the frequency of the waves but not their wave numbers, because they are only related to the time change of the waves. In order to deduce some observational results from wave numbers, we need to have information on the spatial elongation of the waves. One candidate of interest is the shape of the waves in the pulsar timing array (PTA) observations \cite{ipta,pta}. Hence, the following discussion may be relevant observationally since the PTA observations are now publishing their findings, and the wave numbers may be observationally relevant thanks to these experiments.

For tensorial waves, from (\ref{efffreqtn}), we obtain, by remembering $ \nu=k $, that
\begin{equation}
	\frac{\Delta k}{k}\equiv\frac{k-k_{eff}}{k}=H Z/2. \label{tenwaveshift}
\end{equation}

Hence,  if we could have some experimental data regarding parameters, for example $k, k_{eff}$ and $H$ in these relations, we can deduce the remaining parameter, $Z$ or vice versa.
For scalar waves,  from (\ref{keffsc}) we have
\begin{equation}\label{scwaveshift}
	\frac{\Delta \tilde k}{\tilde k}\equiv \frac{\tilde k- \tilde k_{eff}}{\tilde k}=\frac{1}{\sqrt{1-\frac{m_s^2 }{\omega^2}}} \frac{H Z}{2}. 
\end{equation} 
Hence, knowing $\tilde k, \tilde k_{eff}$, $H$ and $Z$ from observations and the tensorial waves,  we can in principle again determine the mass of the scalar field. As we have mentioned before, the mass of the scalar field inversely effects the frequency of the waves and wave number,  reducing the frequency shift and enhancing shift of the wave number due to cosmological expansion of the universe. 

\subsection{Effective frequency to wave number ratio}
An interesting observation for these waves comes from the ratio of equations (\ref{tenfreqshift}) and (\ref{tenwaveshift}) by eliminating $HZ$. This gives the following observation
\begin{equation}
	\frac{\Delta \nu }{\Delta k}=2.
\end{equation}
Hence, for tensorial waves, the frequency shift is twice the wave number shift, a result valid only in the order of approximations used in this work. In other words, the change in the frequency is twice the change in the wavelength.

For scalar waves,  again eliminating $HZ$ from equations (\ref{scfreqshift}) and (\ref{scwaveshift}),  we obtain
\begin{equation}
	\frac{\Delta \omega}{\Delta \tilde k}=2 \frac{\tilde k}{\omega}=2 \sqrt{1-\frac{m_s^2}{\omega^2}}.
\end{equation}	
Since $\tilde k<\omega$ for a massive scalar field, the ratio of frequency shift to wavelength shift is less than 2, which is the corresponding ratio for tensorial waves. If these ratios can be somehow observed, it may be possible to determine the mass of the scalar field, and constrain the parameters therein, defined in eq (\ref{ms}).

\subsection{Detector arrival time difference}
In \cite{Schumacher}, a general framework for GWs having different polarization speeds was investigated, and a simple discussion is given in its appendix,  relating polarization speeds and detector arrival times. Let $t_n$ be the arrival time of the nth polarization mode of the GW from the source to the observer, then 
\begin{equation}
	t_n=\frac{R}{c_n},
\end{equation}
where $R$ is the distance of the source to the detector, and $c_n$ is the speed of the polarization mode.
The differences in arrival times for the two polarizations $a$ and $b$, can be given by \cite{Schumacher}:
\begin{equation}
	\Delta t_{ab}=t_a-t_b =\frac{R}{c_a}-\frac{R}{c_b}.
\end{equation} 
Let $c_{a}$ and $c_b$ be the speed of tensorial and scalar waves with $c_a=c$ and $c_b=v$, respectively.
From this equation, one can obtain an expression for the speed of the massive scalar wave that must have caused a time difference $\Delta t_{ab}$ as \cite{Schumacher}:
\begin{equation}\label{speedtimedif}
	v=\sqrt{1-\frac{m_s^2}{\omega^2}}=\left( \frac{1}{c}-\frac{\Delta t_{ab}}{R} \right)^{-1}.
\end{equation}
Using same figures with \cite{Schumacher} for a source with distance $R=100\, \mbox{Mpc}\approx 10^{24}\, \mbox{m}$ for a time difference 2048 seconds, the speed of massive scalar modes should satisfy $v\ge c\left[1+\mathcal{O}{(10^{-13})}\right],$ which is too small to observe. We may need much longer observational time to observe the possible speed difference of scalar modes. For example, using again the example of \cite{Schumacher} for ten observational years or $3\times 10^8$ seconds,  we see  $v\approx c\left[1+\mathcal{O}{(10^{-8})}\right] $. These results show that even if the propagation speed has a very small difference from the speed of light, the scalar wave can be missed to be observed by the observational window the detector operates, even if it is capable of detecting them.  The equation (\ref{speedtimedif}) also relates the mass of the scalar field to the arrival time difference of scalar waves compared to tensorial waves from the same source. Indeed, it can also be written, using the explicit dependence of the speed of massive scalar waves, as 
\begin{equation}
	m_s=\omega \sqrt{ 1- \left( \frac{1}{c}-\frac{\Delta t_{ab}}{R} \right)^{-2} }.
\end{equation}	
Imagine if the distance of the source, the frequency of scalar waves, and the time arrival difference are spotted in a future observation, then one can find the mass of the scalar field from this observation. This mass value can then constrain the theories admitting massive scalar fields, such as massive BD theory or Horndeski theory, namely, the parameters given in equation (\ref{ms}). 

\section{A Comparison of different implementations of cosmological constant in Horndeski theory}

Our aim in this paper was to determine and analyze GW solutions in the constant curvature background  originated from the minimum of the potential, rather than an asymptotically flat one, as is usually done in the literature. This can be done in the Horndeski theory in different ways, as we will discuss below. Up to this point, we have done this for an arbitrary potential whose minimum plays the role of a cosmological constant. One can call this approach the massive Horndeski theory since the potential produces an effective mass to the scalar field. One way of obtaining this particular theory is to have a potential term $U(\phi)$ in the term $L_2$  in the Horndeski action in (\ref{Hornaction},\ref{Hornactions}),  which can play both the role of an effective cosmological term together with bringing an effective mass to the scalar field. This is the approach we have followed throughout sections II-IV.  In section II, we have presented the linearized equations, and their solutions in section III, and some physical applications of that implementation in section IV. As an example of this approach, we also discuss a particular quadratic potential in the following discussion. We now discuss other ways of achieving an asymptotically de-Sitter background and also their properties by comparing with the massive Horndeski theory that we have already obtained up to now.

\begin{table}[h!]
	\caption{Linearized field equations for various choices to implement cosmological constant in  Horndeski theory}
	\label{table1}
	\begin{tabular}{ | c |c |c |c |}
		\hline\hline
		
		Generic terms &  Conditions  & Tensor equations &  Scalar Equation  \\  \hline \hline
		\multicolumn{4}{|c|}{Massive Horndeski, Arbitrary potential } \\ \hline
		$K(0,\phi)=- U(\phi)$ &  $ K(0)\neq 0, K_{,\phi}(0)=0 $  & $\Box \theta_{\mu\nu}=- \frac{K(0)}{G_4(0)}\eta_{\mu\nu}$, & $(\Box -m_s^2)\varphi=2 K(0)\chi  $	\\ \hline\hline 	\multicolumn{4}{|c|}{ Quadratic   potential  } \\ \hline
		$K(0,\phi)=-\frac{1}{2} m^2 (\phi-\phi_0)^2-V_0$ &  $  K_{,\phi}(0)=0 $  & $\Box \theta_{\mu\nu}=\frac{V_0}{G_4(0)}\eta_{\mu\nu}$, & $(\Box -m_s^2)\varphi=-2 V_0\,\chi  $	\\ \hline \hline  
		\multicolumn{4}{|c|}{ 	Linear potential} \\ \hline
		$K(0,\phi)=-2\Lambda_1 \phi$   & $\Lambda_1=\mbox{constant}$ &  $\Box \theta_{\mu\nu}=\frac{2 \Lambda_1 \phi_0}{G_4(0)}\eta_{\mu\nu}$ & $\Box \varphi=\left(2\frac{G_4(0)}{G_{4,\phi}(0)}-4 \phi_0\right)\chi \Lambda_1  $ \\ \hline \hline	\multicolumn{4}{|c|}{ Vacuum energy } \\ \hline
		$T_{\mu\nu} = -\Lambda g_{\mu\nu}$ or $K(0,\phi)=-2\Lambda$  & &  $\Box \theta_{\mu\nu}= \frac{2 \Lambda }{G_4(0)}\eta_{\mu\nu}$ &  $\Box \varphi=-4\chi \Lambda  $ \\
		
		\hline
	\end{tabular}
	
\end{table}

Another way to implement an  Asymptotically dS background is to have a linear potential, which give rise to a cosmological constant in the equations but leads to a massless theory. A special case of this theory is the BD$\Lambda$ theory \cite{Lorenz-Petzold,Endo,Kim}, whose GW solutions were discussed in \cite{Ozer1}. 

Yet another way is to include a background vacuum energy density as a source, as done in GR. In this case, we suppose that there is a possible vacuum energy density filling the spacetime with the energy-momentum tensor
\begin{equation}
	T_{\mu\nu}=-\Lambda\, g_{\mu\nu},
\end{equation}
with the trace  $T=-4 \Lambda$. Instead of adding a vacuum energy-momentum tensor, we can also achieve this result by adding an appropriately chosen constant term in $ L_2$ in (\ref{Hornaction}) by replacing that with the potential term in massive Horndeski theory, rendering it a massless theory. 

In order to see the similarities and differences of these approaches to have an  Asymptotically  dS spacetime in the presence of linearized GWs, we present the results with the help of several tables. In Table (\ref{table1}), various choices to implement dS asymptotics, choice of the potential or energy momentum tensor, the necessary conditions, corresponding tensor and scalar field equations were listed for massive Horndeski theory with a generic potential or a quadratic potential, Horndeski with a linear potential or vacuum energy density. 

Our aim is to obtain and compare the effects of all these alternative terms, which actually give the theory a dS background. Hence, we choose to include these terms in the formulas to obtain a general solution that includes both terms economically.  We will then compare the results of those cases.  Note that all these terms lead to the same theory,  GR$\Lambda$, in the GR limit. However, their local and or global properties can be very different, as in the special case of BD theory as discussed in various works \cite{Ozer,Barrow}. Hence, it is worthwhile to study the properties of all varieties in the present discussion. We present the results via some tables in this section.

\begin{table}[h!]
	\caption{Solutions of linearized field equations for various choices to implement cosmological constant in  Horndeski theory} \label{table2}
	\begin{tabular}{|c|c|c|}\hline
		Theory& Tensor solution ($h_{\mu\nu}$)& Scalar field ($\varphi$) \\ \hline
		Massive Horndeski    &	$\frac{K(0)}{18G_4(0)}x_\mu x_\nu+\frac{K(0)}{9G_4(0)}\eta_{\mu\nu}x^2+A_{\mu\nu}\sin kx+B_{\mu\nu}\cos kx-\eta^{\mu\nu}\sigma \varphi,$  & $\frac{
			K(0)\chi r^2}{3}+\varphi_0e^{i\,q \,x}$
		\\ \hline Quadratic potential & $K(0)=-V_0$  &$K(0)=-V_0$
		 \\ \hline
Linear potential	& $K(0)=-2 \Lambda_1 \phi_0$, $m_s^2=0$ &$K(0)=\left(\frac{G_4(0)}{G_{4,\phi}(0)}-2 \phi_0\right)\Lambda_1$ , $m_s^2=0$\\  \hline
Vacuum Energy & $K(0)=-2 \Lambda$, $m_s^2=0$& $K(0)=-2 \Lambda $, $m_s^2=0$\\
\hline
	\end{tabular}
	
	\end{table}

In Table \ref{table2}, the solutions of the scalar and tensorial equations are listed. The metric perturbation tensor $h_{\mu\nu}$ and the scalar field $\varphi$ are presented for massive Horndeski theory in the first row. The other solutions can be obtained if the indicated changes were made in the corresponding rows of the other prescriptions.

\begin{table}[h!]
	\caption{Polarization properties and propagation speeds of gravitational waves for various choices to implement the cosmological constant in  Horndeski theory. Here $+$ and $\times$ correspond to transverse plus and cross polarizations of GR, whereas $b$ corresponds to the transverse scalar breathing mode, and $l$ represents longitudinal scalar mode. }
	\label{table3}
	\begin{tabular}{|c|c|c|}\hline
		Theory& Polarizations& Propagation Speeds \\ \hline
		Massive Horndeski theory  & $+,\times,b,l$  & ($+,\times$ speed of light), ($b,\l$ less than $c$)
		\\ \hline Quadratic potential & $+,\times,b,l$  & ($+,\times$ speed of light), ($b,\l$ less than $c$) \\ \hline
		Linear potential	& $+,\times,b$ & All speed of light\\  \hline
		Vacuum Energy &$+,\times,b$ &  All speed of light\\
		\hline
	\end{tabular}
	
\end{table}

In table \ref{table3}, physical properties of the gravitational waves, such as polarizations and propagation speeds of the tensorial and scalar waves, are presented. For all the cases, we have usual plus and cross tensorial polarizations with the speed of light. For massive Horndeski and quadratic potential cases,  breathing and longitudinal scalar polarizations that move with a speed smaller than the speed of light also exist. For the linear potential and vacuum energy cases, only the breathing scalar mode exists, which propagates at the speed of light. Table \ref{table3} clearly shows that we can have gravitational waves  with different polarization properties in an asymptotically dS spacetime in linearized Horndeski theory.

\begin{table}[h!]
	\caption{Polarization properties and propagation speeds of gravitational waves for various choices of scalar field $G_4$ coupled to Ricci scalar and potential term $K$ in viable Horndeski theories. We assume a minimal coupling between the scalar field and matter-energy Lagrangian. }
	\label{table4}
	\begin{tabular}{|c|c|c|p{4cm}|}\hline
		$G_4(\phi,0),K(\phi,0)$& Polarizations& Propagation Speeds&  Viable Horndeski Theories \\ \hline \hline
		$G_4(\phi,0)=$constant, $K$ arbitrary & $+,\times$  &  All speed of light& GR, Einstein-Scalar, Quintessence/K-essence, Kinetic Gravity Braiding
		\\ \hline $G_4(\phi,0)=f(\phi)$, $K(\phi,0)=0$ & $+,\times,b$  &  All speed of light & Brans-Dicke ($f(\phi)=\phi$) \\ \hline
		$G_4(\phi,0)=f(\phi)$, $K(\phi,0)=\Lambda_1\phi+b$ & $+,\times,b$  & All speed of light & BD with a CC \\
		\hline
		$G_4(\phi,0)=f(\phi)$, $K(\phi,0)=U(\phi)$, $K_{\phi\phi}\neq 0$	& $+,\times,b,l$ & ($+,\times$ with c), ($b,l$ less than c) & BD with a potential, massive BD, $f(R)$ \\  \hline
	\end{tabular}
	
\end{table}

Table \ref{table4} is devoted to a discussion of the properties of the GWs in viable Horndeski theories. The observation GW170817 \cite{Ligomm} severely restricts various modified gravity theories, including Horndeski theory \cite{Ezquiaga}, where some subcases of this theory are eliminated, such as quartic/quintic Galileons, Fab Four, etc. Hence, in Table \ref{table4} we present the propagation and polarization properties of the GWs in the viable Horndeski theories that are compatible with GW170817. The form of $G_4$ and $K$ determines these properties as listed in Table IV.  

\section{Conclusions}

In this paper, we have determined and analyzed linearized GW solutions in the constant curvature background originated from the minimum of the arbitrary potential, rather than the asymptotically flat one usually discussed in the literature for the Horndeski theory.
 Since the scalar field has an effective mass in general for this case, it has a short range, as in massive BD theory and $f(R)$ theories. We find that the background geometry is different then the linearized  version of de Sitter metric, due to the effects of the scalar perturbation in Horndeski theory with an arbitrary potential. Only if the scalar field vanishes, then the background geometry reduce to linearized dS metric.

 After obtaining these solutions, we have discussed the properties of GWs, their propagation, and interaction with the medium in which they propagate by taking into account the background curvature.  We have studied the effects of these waves on test particles and detectors by calculating the geodesic deviation of two test particles initially at rest due to the GW. In GR, the GWs present only two polarization states, the $+$ and $\times$ modes. However, within the framework of massive Horndeski theory, the polarization content of GW is not limited to the two tensor modes of GR due to the presence of additional degrees of freedom \cite{Houa}. As shown in previous analyses, we find that GWs possess three independent polarization modes, two tensor modes corresponding to the usual plus and cross polarizations predicted by GR, and one massive scalar mode, which induces both transverse breathing and longitudinal components \cite{Dicong,Houa}. In the massless limit of the scalar field, the longitudinal component is absent and only the transverse breathing mode remains.
We find that the combined effect of the effective cosmological constant and the background scalar field produces a homogeneous  expansion between the test particles. This behavior is analogous in a similar way to the expansion driven by a cosmological constant.The obtained solutions were in accordance with the sub-theories of Horndeski theory. These waves behave similarly to GWs in massive BD or $f(R)$ theories. Therefore, the solutions and analyses obtained in this theory can be used to test the predictions of different gravity theories at various scales.

The massive case is not the only way to have an asymptotically nonflat dS like background in Horndeski theory. The other ways include a linear potential leading to a massless long-range scalar together with a cosmological constant or a vacuum energy density filling the universe. The differences between massless, linear potential, or vacuum energy cases can be realized in the propagation speeds and polarization properties of these waves.  A detailed comparison of the properties of these alternative approaches was discussed in Section IV via several tables. Only if the potential is an arbitrary function of the scalar field whose second derivative is non-vanishing,  the longitudinal scalar mode is present, and scalar waves propagate with a speed less than the speed of light. Otherwise, only the breathing mode is present, which propagates with the speed of light.  The form of the background solution for this theory also has some numerical differences for different choices of the potential, since the scalar equations and their solutions will be different for each choice of the scalar potential, as we have presented in the last columns of Tables I and II. This observation is also compatible with the previous work \cite{Ozer} for BD theory with different choices to have a cosmological constant.

The effect originated from the minimum of the potential is to cause an accelerated expansion of the nearby test particles, mimicking the effective cosmological constant, causing expansion of the universe. In order to see these effects better, we have transformed the solution into a-FRW-type coordinates. The obtained solution then enables us to analyze the effect of this expansion on the frequency and wave vector of these waves for both tensorial and scalar GWs. The waves are redshifted by cosmological expansion. The wave vector is also redshifted, but for both tensorial and scalar waves, the amount of redshift is different for wave frequency and wave vector. This effect was  realized before for GR \cite{Bernabeu1},  but we have extended this result to general scalar-tensor theories in this work and massless scalar fields \cite{Ozer} for BD theory.  The expressions we have obtained and also their differences for scalar and tensorial waves may open an observational window to determine properties of the  Horndeski theory and its sub-theories, such as the effective Newton constant, minimum of the potential, and mass of the scalar field in future observations.  

We know that current wave observations are in agreement with the predictions of GR, and the waves are transverse and travel at the speed of light. However, according to the results for Horndeski theory and other scalar-tensor theories in the literature, these waves may also have scalar modes. Current GW detectors are insufficient to resolve these scalar modes, even if they exist. However, if scalar waves and their propagation speeds can be observed in other detector systems planned to be built in the future using a network of detectors with different orientations, the presence or mass of scalar waves can be determined, or limits can be put on their mass and speed. Therefore, we think that our results can contribute to the results of future detector systems, especially future space-borne detectors.

There may be several future directions that this work may inspire. For example, application of theories having massless or massive scalar fields with effective cosmological constant to Pulsar Timing Array (PTA) observations, as was done in the GR case in the works \cite{EspriuPTA,Espriu}, or generalizing the solutions by including other cosmic fluids \cite{EspriuRodera,Espriu1} could be a future direction to generalize this work. Another direction could be focusing on local effects and scattering of these waves in Horndeski theory, which we are currently working on, in the presence of a nontrivial scalar potential term.

\section*{Acknowledgement}

This study was supported by Scientific and Technological Research Council of Turkey (TUBITAK) under the Grant Number 122F331. The authors thank to TUBITAK for their support.

\section*{Supplementary information}
There is no supplementary file regarding this paper.
\section*{Data availability Statement}
No data associated in the manuscript.

\bibliography{References}

\end{document}